# Cross-Modal Contrastive Representation Learning for Audio-to-Image Generation

HaeChun Chung, JooYong Shim, Jong-Kook Kim*

Dept. of Electrical Engineering, Korea University

marcomx@korea.ac.kr, shimjoo@korea.ac.kr, jongkook@korea.ac.kr

**Abstract.** Multiple modalities for certain information provide a variety of perspectives on that information, which can improve the understanding of the information. Thus, it may be crucial to generate data of different modality from the existing data to enhance the understanding. In this paper, we investigate the cross-modal audio-to-image generation problem and propose Cross-Modal Contrastive Representation Learning (CMCRL) to extract useful features from audios and use it in the generation phase. Experimental results show that CMCRL enhances quality of images generated than previous research.

## 1  Introduction

Information can be embedded in multiple modalities and this may allow a better understanding of the information. For example, video data are usually represented by audio and image modalities. Even though both modalities have different data, both are interrelated to one another and helps in the understanding of the data. In real world applications, even if multiple modalities are possible for certain information, one modality may be missing and if it can be generated, it will improve the understanding of that information. Because different modalities have different properties it is difficult to translate or generate a certain modality from existing different modality. This paper focuses on the audio-to-image generation problem, where appropriate images are generated from the audio input. The proposed method trains both the audio and image encoders to extract features using a constrastive learning method [1] and the features are inserted as condition to a version of a Generative Adversarial Networks (GANs) [2] for the image generation. This paper presents Cross-Modal Contrastive Representation

* Corresponding Author: Jong-Kook Kim (jongkook@korea.ac.kr)

** Acknowledgement: This research was supported by Basic Science Research Program through the National Research Foundation of Korea(NRF) funded by the Ministry of Education(NRF-2016R1D1A1B04933156).

Learning (CMCRL), which uses the supervised constrastive learning idea [1] to make the latent features of different modalities closer, as shown in Fig 1(b). The result is that audio features similar to image features are extracted and is inserted to a version of a GANs such that higher quality images than previous research are generated.

The rest of this paper is organized as follows: Section 2 summarizes related research. The proposed method is introduced in Section 3. Section 4 shows and analyzes the results and then the last section concludes this paper.

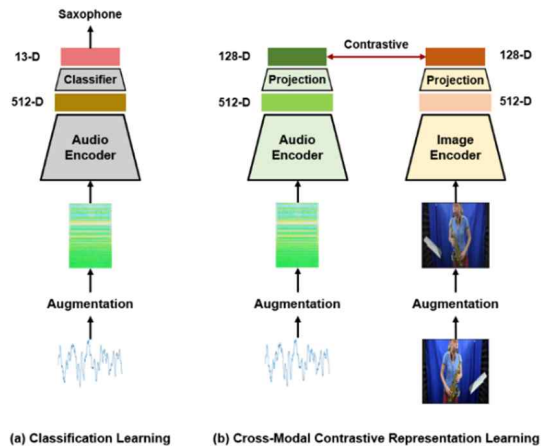

**Fig. 1.** The model of (a) Classification Learning using only audios and (b) Cross-Modal Contrastive Representation Learning using both audios and images.

## 2 Related Works

[3] was the first paper that conducted cross-modal audio-visual generation using the Sub-URMP dataset. They used pre-trained encoders that was used for classification to extract audio features for the condition in Conditional Generative Adversarial Networks (CGANs). Our method is similar to the structure of [3], but the internal details of the structure are different. First, CGANs is replaced with a Self-Attention Generative Adversarial Networks (SAGAN) [2]. Second, the classification learning that used only audios is replaced with Cross-Modal Contrastive Representation Learning (CMCRL), which trains to make the features of both audios and images to be similar.

[4] proposes a Cross-Modal Cycle Generative Adversarial Networks (CMCGAN) that enables end-to-end training by integrating visual-audio mutual generation into a common framework using the encoder-decoder cycle architecture. The paper [5] introduces self-attention and cross-attention modules that are included in the CMCGAN. The research in [6] proposed a model using a cascade coarse-to-fine generation strategy by introducing a residue module between two generators, and achieved state-of-the-art performance. [6] uses class labels to generate images. In this paper, CMCRL is used to reduce the gap between different modalities and this affects the generation model to generate higher quality images and accuracy performance comparable to [6] without including labels.

## 3  Proposed Method

**3.1 Cross-Modal Contrastive Representation Learning**

In this paper, to extract audio features, the audio encoder are trained using Cross-Modal Contrastive Representation Learning (CMCRL) with image encoder, shown in Fig 1(b), instead of classification learning only using audios, shown in Fig 1(a). These audio features that extracted pretrained audio encoder are inserted to the image generation phase that uses a Generative Adversarial Networks (GANs).

CMCRL is similar to the structure of [1], except that encoders and projection networks for the different modalities (audios and images) do not share parameters. Given a mini-batch N of audio/image/label pairs, $\{a_l, v_l, y_l\}_{l=1...N}$, augmentations are applied to audios and images and depicted as $\{\tilde{a}_l, \tilde{v}_l, y_l\}_{l=1...N}$. Audio files are sampled at 44,100 Hz and are augmented using fade in/out and time masking that showed the best performance in [7]. Images are resized to 256×256 and then random crop, horizontal flip, and color distortion augmentations are applied. In the case of audios, after augmentations is applied, a pre-processing step is applied to convert the audio signals into 128×44 size of time-frequency audio features composed of Magnitude STFT, Phase STFT, and Mel-Spectrogram that are used in [7]. Augmented time-frequency audio features and images are forward propagated through the audio encoder network $f_a(\cdot; \theta_a)$ and image encoder network $f_v(\cdot; \theta_v)$, respectively, to extract 512-dimensional features of audios and images. Finally, these features are propagated through an audio projection network $g_a(\cdot; \phi_a)$ and an image projection network $g_v(\cdot; \phi_v)$, respectively, to extract 128-dimensional features of audios and images.

$$z_l^a = g_a(f_a(\tilde{a}_l; \theta_a); \phi_a) \qquad (1)$$

$$z_l^v = g_v(f_v(\tilde{v}_l; \theta_v); \phi_v) \qquad (2)$$

Audio feature representations $z_l^a$ and image feature representations $z_l^v$ are normalized to be placed in the 128-dimensional unit sphere, which enables the calculation of the distance between feature vectors using the inner product in the embedding space of the projection. Then, Audio feature representations $z_l^a$ and image feature representations $z_l^v$ are concatenated to make a total of 2N features $z_i$, $i \in I \equiv \{1 \dots 2N\}$. The CMCRL loss is defined as:

$$\mathcal{L}_{AV} = -\sum_{i \in I} \frac{1}{|P(i)|} \sum_{p \in P(i)} \log \frac{\exp(z_i \cdot z_p / \tau)}{\sum_{t \in T(i)} \exp(z_i \cdot z_t / \tau)} \qquad (3)$$

Here, $T(i) \equiv I \setminus i$, $P(i) \equiv \{p \in T(i) : y_p = y_i\}$, $\cdot$ denotes the inner (dot) product, and $\tau$ is the temperature parameter. As mentioned in [1], this loss causes the anchor to pull together the positive samples belonging to the same class as the anchor in the embedding space, and to push apart from the negative samples belonging to the other class. This encourages close clustering of the feature representations of all entries in the same class, which results in a more robust clustering of the embedding space. Furthermore, because audio and image features are concatenated, this loss causes the distances of audio-audio, image-image, and audio-image features belonging to the same class to cluster near each other. The CMCRL method is compared to the classification learning method to show the effectiveness in the classification task. The results of CMCRL is inserted into a GANs to generate images from audio data.

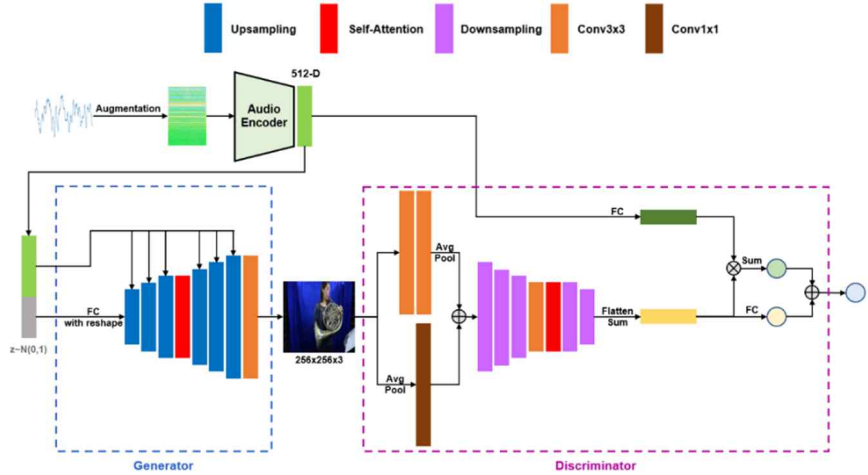

**Fig. 2.** The overall model architecture of the proposed method. Features that are extracted from pretrained audio encoders using the Cross-Modal Contrastive Representation Learning (CMCRL) method are utilized as conditions for the GANs.

### 3.2 Audio-to-Image Generation

The pretrained audio encoder using the CMCRL and Self-Attention Generative Adversarial Networks (SAGAN) [2] is used to generate an output image that matches the audio input, as shown in Fig 2. SAGAN introduced a self-attention mechanism to the GANs, which complements convolution by capturing long-range, global-level dependencies. In addition, spectral normalization was applied to not only the discriminator but also the generator to stabilize the training, and the two time-scale update rule was applied to speed up the training of the regularized discriminator.

The proposed model architecture is similar to the SAGAN except for the condition part. The SAGAN takes random noise and embedded class label vector as conditions to generate an image for the particular class. We replaced embedded class labels with audio features $c = f_a(\tilde{a}_l; \theta_a)$ extracted from the pretrained audio encoder discarding the projection part as shown in [1]. In Audio-to-Image Generation, the same audio preprocessing settings determined from CMCRL are applied, and images are only resized to 256×256 and no augmentation is applied. The hinge version of the adversarial loss is minimized to train the SAGAN, and thus the loss is defined as follows:

$$\mathcal{L}_D = -\mathbb{E}_{c \sim p_{f_a}, v \sim p_{image}}[\min(0, -1 + D(v, c))]$$
$$- \mathbb{E}_{z \sim p_z, c \sim p_{f_a}}[\min(0, -1 - D(G(z, c), c))], \qquad (4)$$
$$\mathcal{L}_G = -\mathbb{E}_{z \sim p_z, c \sim p_{f_a}} D(G(z, c), c)$$

## 4 Results

**4.1 Experiment Setup**

The performance of model was evaluated using the Sub-URMP dataset [3] that consists of 0.5 second long audio files and image files for 13 types of instruments. Audios or images are extracted to 512-dimensional feature representations using the ResNet-18 as the base encoder network, and additionally projected into the 512-512-128-dimensional latent space using a two-layer MLP projection head. All network for Cross-Modal Contrastive Representation Learning (CMCRL) are trained from scratch using Stochastic Gradient Descent (SGD) using the initial learning rate of 0.05 that decays by 0.1 at 500, 700, 800 epochs, weight decay of $1e^{-4}$, momentum of 0.9, and the batch size of 1024 on V100 GPU for a total of 1000 epochs.

The network structure of the generator and discriminator is similar to [2], but the networks are expanded to generate 256×256 images. Generator and discriminator are trained, but audio encoder network is frozen. Adaptive Moment Estimation (Adam) optimizer using $\beta_1 = 0.5$ and $\beta_2 = 0.999$ is used for the generator and discriminator training on V100 GPU and learning rate for the discriminator is 0.0004 and the learning rate for the generator is 0.0001, and the batch size is 32.

**Table 1.** Classification accuracy of CMCRL and classification learning on Sub-URMP.

| Method | Test Accuracy | |
|---|---|---|
| | Audio | Image |
| Classification Learning | 90.22 | 95.93 |
| CMCRL | 93.49 | 99.99 |

**4.2 Results on Classification**

**Classification Accuracy** To measure the classification performance, after CMCRL, linear classifiers are respectively trained on top of the frozen audio and image base encoders, ResNet-18, without projection layer. In the case of Classification Learning, ResNet-18 and linear classifier were trained as a whole for audio and image, respectively. This task is to test whether CMCRL extracts suitable feature representations for the classification. The classification accuracy is compared to audio and image encoders trained using classification learning. As shown in Table 1, the encoder pretrained using CMCRL outperforms the classification learning for both audio and image, which demonstrates that the feature representation learning provides stronger clustering as mentioned in section 3.

**Table 2.** IS and FID of generated images using various different methods.

| Method | IS | FID |
|---|---|---|
| S2I-C | 2.315 | 252.66 |
| CMCGAN | 2.883 | 215.92 |
| CAR-GAN | 3.818 | 207.37 |
| Classification Learning + SAGAN | 2.757 | 178.89 |
| CMCRL + SAGAN | 5.288 | 107.26 |

**Table 3.** The classification accuracy for generated images based on different models.

| Method | Accuracy | |
| --- | --- | --- |
| | Train | Test |
| S2I-C | 0.8737 | 0.7556 |
| CMCGAN | 0.9105 | 0.7661 |
| CAR-GAN | 0.9954 | 0.9068 |
| Classification Learning + SAGAN | 0.9677 | 08448 |
| CMCRL + SAGAN | 1.0000 | 0.8907 |

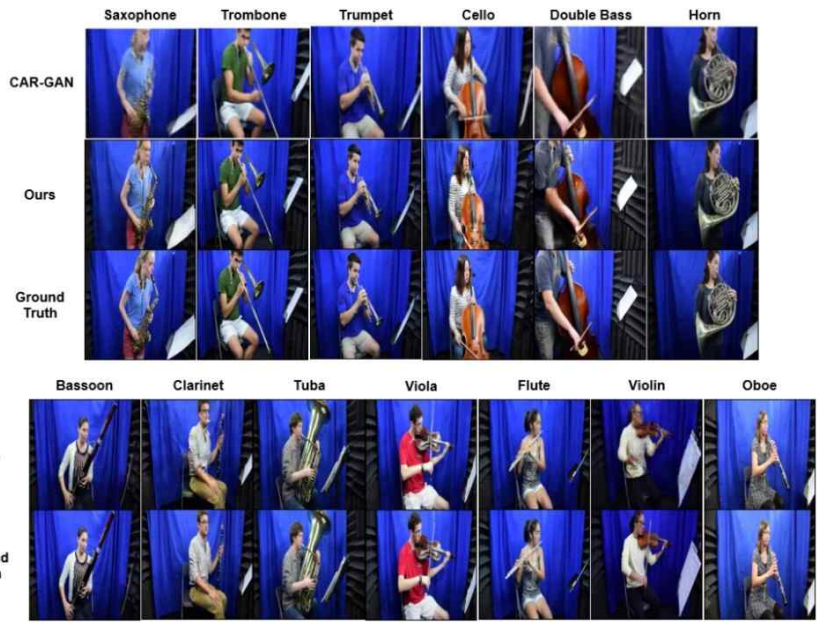

**Fig. 3.** Generated Images of CAR-GAN and proposed method using the Sub-URMP dataset compared to ground truth iamges.

### 4.3 Results on Audio-to-Image Generation

The proposed method is compared to the existing methods [3, 4, 5, 6] using the Sub-URMP dataset. Since the codes of previous works is not available, the experimental results of previous works are the same as those presented in the research. The proposed method and classification learning using Self-Attention Generative Adversarial Networks (SAGAN) are compared to the previous audio to image generation methods.

**Quantitative Evaluation** As metrics for quantitatively evaluating the generated images, Fréchet Inception Distance (FID) and Inception Score (IS) are used, which are widely used metrics to evaluate generated images, and the results are shown in Table 2. The results show that the proposed method significantly outperforms existing methods for both FID and IS and it can be inferred that extracting better feature representations may be the key to generating high quality images.

The classification accuracy of generated images is also compared, which is the metric used in previous works [3, 4, 6], and the results are shown in Table 3. The results show that the proposed method shows the best classification accuracy that is comparable to CAR-GAN [6]. The CAR-GAN scheme uses explicit labels to generate images, but the proposed method does not require labels and only relies on audio features extracted from audio encoder pretrained using CMCRL. One clear advantage of the proposed method is that it is able to generate high quality images even when there are no labels

**Qualitative Evaluation** The generated images from the proposed method for all 13 classes are compared to the ground truth and five images from CAR-GAN is shown in Fig 3. The limited number of images is due to the lack of usable code and thus images are directly copied from the previous paper. Only five are available from the paper. For most classes, high quality images very similar to the ground truth images are generated from our method. In addition, compared to generated images from previous method, the proposed method presents higher quality images.

## 5   Conclusions

In this paper, we present Cross-Modal Contrastive Representation Learning (CMCRL) that learns audio features similar to image features. CMCRL effectively reduces the gap between different modalities and as a result provides useful audio features that is inserted into the Self-Attention Generative Adversarial Networks (SAGAN) for audio-to-image generation. The proposed method is compared to previous audio-to-image generation methods as well as audio-to-image generation that uses classification learning to extract audio features. These experimental results demonstrate that our method generates higher quality images compared to the existing methods.